\documentclass[epj,twocolumn]{webofc}
\usepackage[varg]{txfonts}   % Web of Conferences font

\usepackage{epstopdf}
\def\bea{\begin{eqnarray}}
\def\eea{\end{eqnarray}}

\def\pp{\mbox{$p$-$p$}}
\def\pa{\mbox{$p$-A}}

\def\auau{\mbox{Au-Au}}

\def\aa{\mbox{A-A}}

\def\ee{\mbox{$e^+$-$e^-$}}
\def\ppbar{\mbox{$p$-$\bar p$}}

\def\pt{$p_t$}
\def\yt{$y_t$}
\def\nch{$n_{ch}$}

\woctitle{XLIV International Symposium on Multiparticle Dynamics}
\begin{document}
\title{QCD prediction of jet structure in 2D trigger-associated momentum correlations and implications for multiple parton interactions}

\author{Thomas A.\ Trainor\inst{1}
%\fnsep
%\thanks{\email{Mail address for first author}} 
        % etc.
}

\institute{CENPA 354290 University of Washington, Seattle, Washington, USA}

\abstract{%
The expression ``multiple parton interactions'' (MPI) denotes a conjectured QCD mechanism representing contributions from {\em secondary} (semi)hard parton scattering to the transverse azimuth region (TR) of jet-triggered \pp\ collisions. MPI is an object of underlying-event (UE) studies that consider variation of TR \nch\ or \pt\ yields relative to a trigger condition (leading hadron or jet \pt). An alternative approach is 2D trigger-associated (TA) correlations on hadron transverse momentum \pt\ or rapidity \yt\ in which all hadrons from all \pp\ events are included. Based on a two-component (soft+hard) model (TCM) of TA correlations a jet-related TA hard component is isolated. Contributions to the hard component from the triggered dijet and from secondary dijets (MPI) can be distinguished, including their azimuth dependence relative to the trigger direction. Measured \ee\ and \ppbar\ fragmentation functions and a minimum-bias jet spectrum from 200 GeV \ppbar\ collisions are convoluted to predict the 2D hard component of TA correlations as a function of \pp\ collision multiplicity. The agreement between QCD predictions and TA correlation data is quantitative, confirming a dijet interpretation for the TCM hard component. The TA azimuth dependence is inconsistent with conventional UE assumptions.
}
\maketitle
\section{Introduction}   \label{intro}

%Problem, with its history: jets vs flows, assumptions, two paradigms, how to improve model testing with more concise analysis including more features that depend on one model or the other. 

In a high-energy physics (HEP) context dijet production has long been accepted as an important mechanism for hadron formation by parton fragmentation~\cite{hepjets,rick1,rick2}. But in a heavy-ion context ``freezeout'' from a flowing bulk medium is assumed to be the nearly-exclusive hadron production mechanism~\cite{bec,perfliq1,perfliq2}, and there are claims that bulk-medium collectivity may  even play a role in  \pp, $p$-A and $d$-A  systems~\cite{bron}. Previously-accepted contributions from minimum-bias (MB) dijets to high energy collisions are displaced by the claimed presence of strong collective motion (flows) in a thermalized bulk medium or quark-gluon plasma to explain spectrum and correlation structure.
 
Such claims are based in part on the {\it a priori} assumption that all hadrons with transverse momentum $p_t < 2$ GeV/c emerge from a thermalized bulk medium~\cite{2gev}. But that interval includes more than 90\% of  minimum-bias (MB) jet fragments, according to jet measurements and QCD predictions~\cite{eeprd,hardspec,fragevo,jetspec}. Analysis of spectra and correlations does appear to confirm a dominant role for dijet production at low \pt\ in all collision systems~\cite{ppprd,hardspec,fragevo,porter2,porter3,axialci,anomalous}. But we can further extend the QCD description of MB dijet manifestations at low \pt, at least in \pp\ collisions. 

In previous studies a jet-related {\em hard  component} was  isolated from the \pt\ spectrum of 200 GeV \pp\ collisions by means of its charge multiplicity \nch\ dependence, leading to a two-component (soft+hard) spectrum model (TCM)~\cite{ppprd}. 
%A similar differential analysis of \auau\ spectra revealed that the spectrum hard component persists for all centralities, although the form changes quantitatively in more-central collisions~\cite{hardspec}. 
The spectrum hard component for \pp\ spectra has been described quantitatively by a pQCD calculation~\cite{fragevo} based on a MB jet (hard-scattered parton) spectrum~\cite{jetspec2} and measured parton fragmentation functions (FFs)~\cite{eeprd}. Those results establish that the spectrum and angular-correlation hard components in \pp\ collisions are jet related~\cite{hardspec,fragevo},  

The present study extends that program with a method to predict 2D trigger-associated (TA) hadron correlations arising from dijets produced in high-energy \pp\ collisions based on measured FFs and a large-angle-scattered parton spectrum~\cite{duncan,pptrig,jetcorr}. Identification of unique triggered-dijet contributions to TA correlations in \pp\ collisions may then probe the phenomenon of multiple parton interactions (MPI) and test claims of bulk collectivity in \mbox{$p$-A}, $d$-A and  heavy ion (\aa) collisions at the Relativistic Heavy-Ion Collider (RHIC) and the Large Hadron Collider (LHC).

In this paper we define 2D trigger-associated (TA) correlations, derive a TA two-component model (TCM)  and extract a TA hard component (HC) that may represent dijet fragments. We then derive a system to predict the TA HC via a measured MB jet spectrum and parton FFs. Comparisons with measured TA hard components serve to identify kinematic limits on dijet formation in p-p collisions. We isolate {\em triggered} dijets from {\em secondary} dijets (MPI) and test underlying-event (UE) assumptions.

%%%%%%%%%%%%%%%%%%
\section{p-p spectra and dijet production}  \label{ppspec}

The TCM for 2D TA correlations is based on the TCM for 1D \yt\ spectra from \pp\ collisions which we briefly summarize here. The \pp\ spectrum TCM is derived from the \nch\ dependence of \pt\ or transverse rapidity $y_t \equiv \ln[(m_t + p_t)/m_\pi]$ spectra. A ``soft'' component $S(y_t)$ with fixed shape on \yt\ is observed to scale with multiplicity approximately proportional to \nch. A ``hard'' component $H(y_t)$, also with fixed shape on \yt, is observed to scale approximately as $n_{ch}^2$. The soft+hard terminology was adopted when the two components were later associated with specific physical processes via physical model comparisons.

\begin{figure}
\centering
\includegraphics[width=.45\columnwidth]{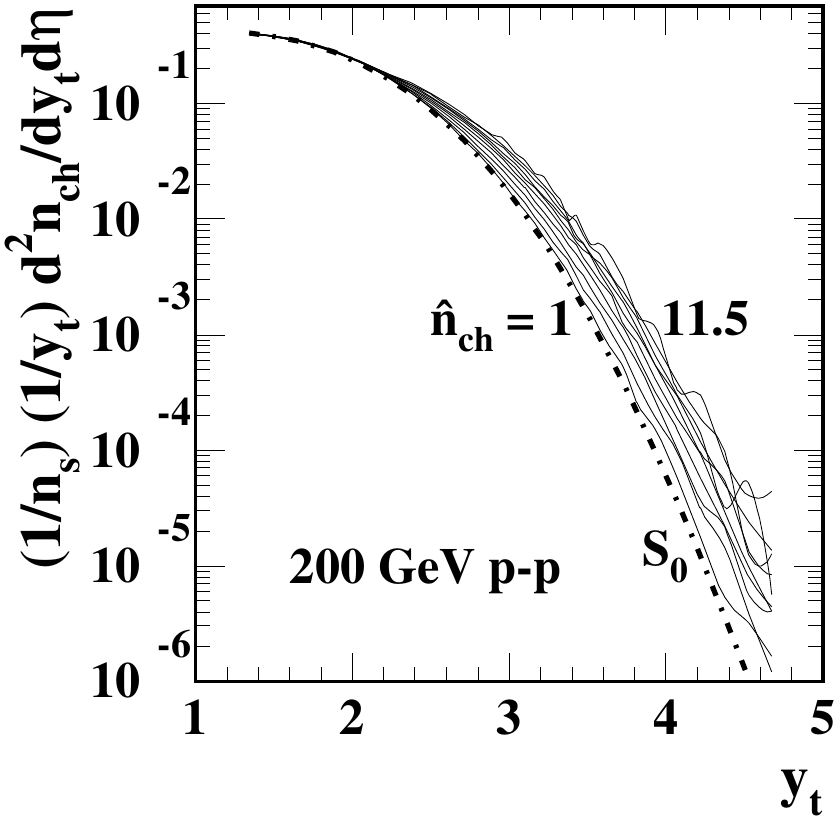}
\put(-30,85){\bf (a)}
\includegraphics[width=.45\columnwidth]{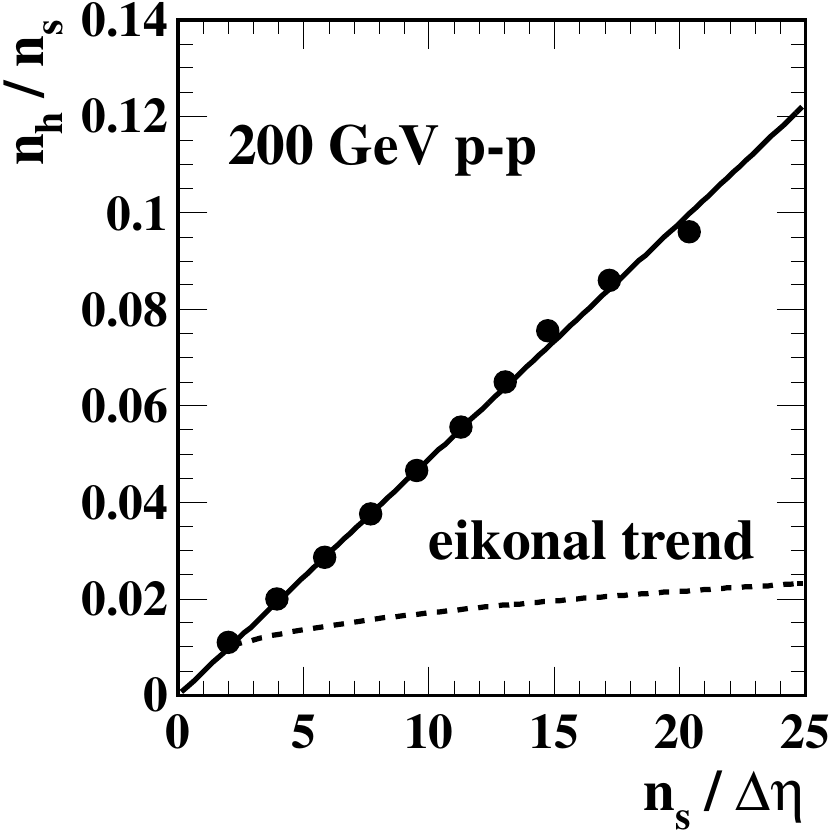}
\put(-30,85){\bf (b)}\\
\includegraphics[width=.45\columnwidth]{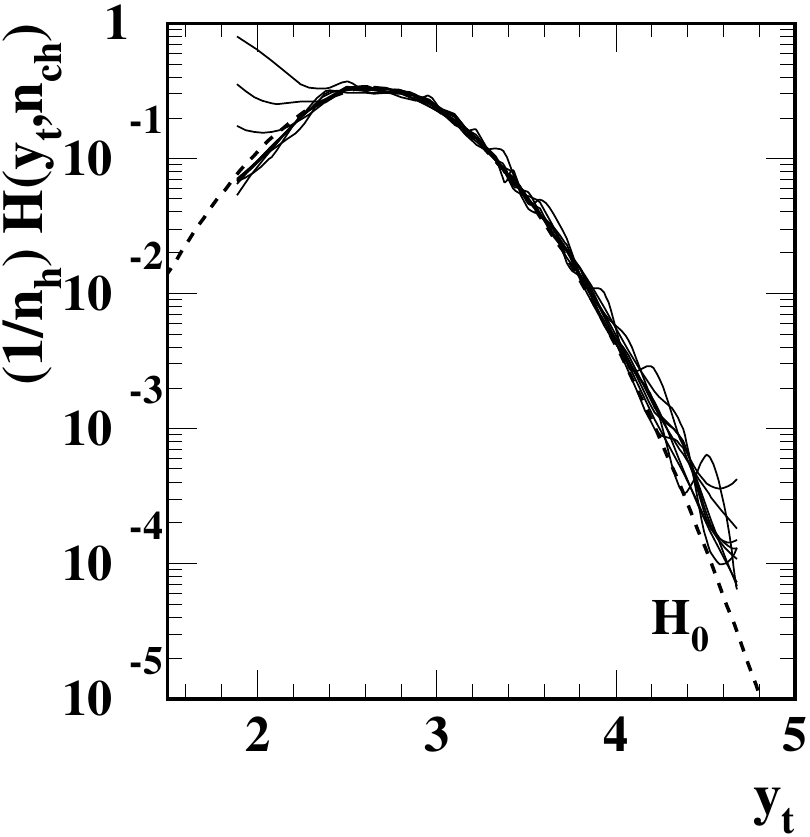}
\put(-30,85){\bf (c)}
\includegraphics[width=.465\columnwidth]{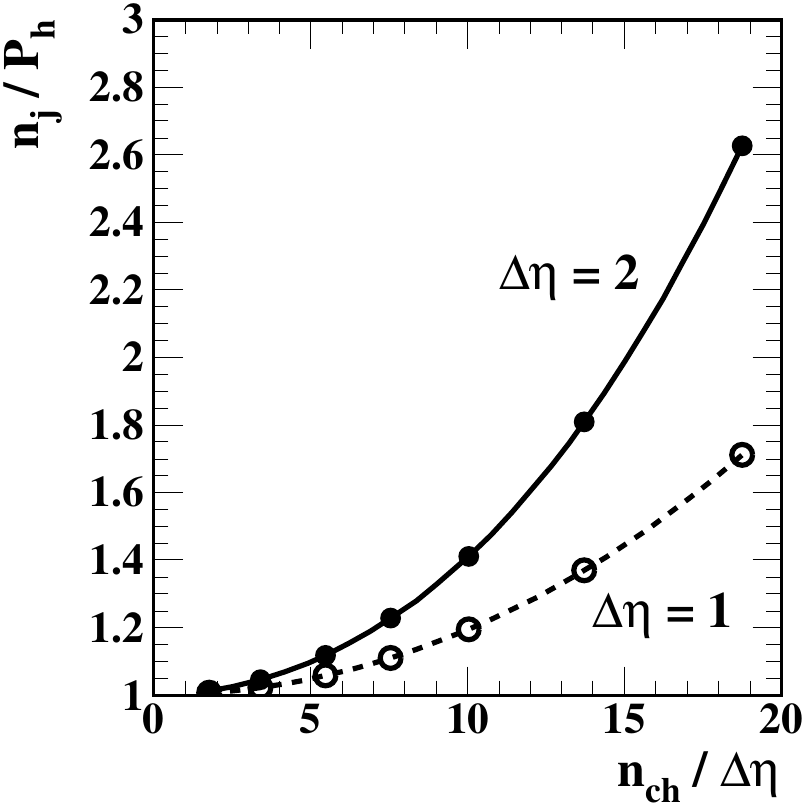}
\put(-30,85){\bf (d)}
\caption{
(a) $y_t$ spectra for eleven multiplicity classes from 200 GeV \pp\ collisions.
(b) Ratio of hard/soft multiplicities $n_h/n_s$ vs the soft multiplicity $n_s$.
(c) Spectrum hard components in the form $H(y_t)/n_h$ from the spectra in panel (a).
(d) Number of dijets per hard \pp\ event $n_j / P_h$ vs total multiplicity density $n_{ch}/\Delta \eta$.
}
\label{ppspec}       
\end{figure}

Figure~\ref{ppspec} (a) shows spectra for eleven multiplicity classes (within one unit of $\eta$) normalized by soft-component multiplicity $n_s$ derived (after the fact) from the hard-component systematics. (The parameter $\hat n_{ch}$ is the uncorrected observed multiplicity, about 50\% of the corrected multiplicity.) The fixed soft-component model $S_0(y_t)$ is represented by the dash-dotted curve. Subtracting that curve from the normalized spectra results in hard components $H(y_t)/n_s$ with integrals $n_h / n_s$ observed to be proportional to $n_s$ as shown in panel (b). The ratio $H/n_h$ then has a fixed shape and amplitude as shown in panel (c), with unit-normal model function $H_0(y_t)$ (dashed curve) approximately Gaussian on \yt\ near the distribution mode. 

Based on a dijet hypothesis we have $\rho_h(n_{s}) = n_h / \Delta \eta = \alpha \rho_s^2 \equiv f(n_{s}) \epsilon(\Delta \eta) 2 \bar n_{ch,j}$, where $\alpha \approx 0.0055$ is inferred from the second panel, $f(n_s)$ is the dijet frequency per event per unit $\eta$, $2 \bar n_{ch,j}$ is the MB dijet fragment multiplicity into $4\pi$ ($\approx 3.5$ for 200 GeV \ppbar) and $\epsilon(\Delta \eta) \in [0.5,1]$ is the fraction of a triggered dijet that appears in the angular acceptance. Given those systematics we can estimate the probability of soft (no dijet in the acceptance) and hard (at least on dijet) event types as $P_s(n_s) = \exp[- \Delta \eta f(n_s)]$ and $P_h = 1 - P_s$. The number of dijets per hard event $\Delta \eta f / P_h$ ($\geq 1$) for two $\Delta \eta$ values is plotted in panel (d).

%%%%%%%%%%%%%%%%%%
\section{Trigger-associated (TA) TCM}  \label{tatcm}

2D TA correlations are constructed as follows: From each \pp\ event with \nch\ hadrons in acceptance $\Delta \eta$ select the highest-\pt\ hadron as the “trigger;” the $n_{ch} - 1$ other hadrons are “associated.” Form all trigger-associated pairs (excluding self pairs) with no \pt\ cuts (resulting in MB TA correlations, no hadrons excluded), the joint pair density denoted by $F(y_{ta},y_{tt}) = T(y_{tt}) A(y_{ta}|y_{tt})$. Subtract calculated TCM TA soft components to obtain conditional associated TA data hard component $A_{hh}(y_{ta}|y_{tt} )$. For UE/MPI studies determine the associated hard-component azimuth dependence relative to the trigger direction.

The TA TCM is based on the assumption that there are no correlations; each hadron represents an independent sample from the fixed single-particle (SP) spectrum. The trigger-particle spectrum $T(y_{tt})$  is modeled by the compound probability that a hadron appears at $y_{tt}$ (the normalized SP hadron spectrum) and no other hadron in that event has a larger \yt, represented by Poisson {\em void probability} $G(y_{tt}) = \exp[ - n_{\Sigma}(y_{tt})]$, where $n_{\Sigma}(y_{tt})$ is the spectrum integral above $y_{tt}$.

\begin{figure}
\centering
\includegraphics[width=.485\columnwidth]{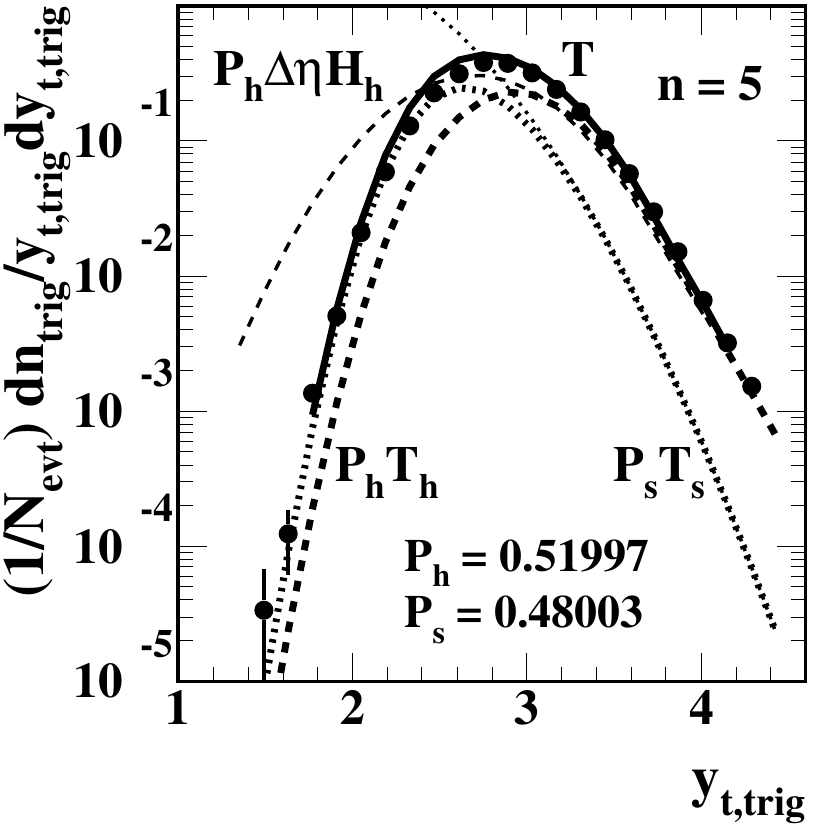}
\put(-82,90){\bf (a)}
\includegraphics[width=.47\columnwidth]{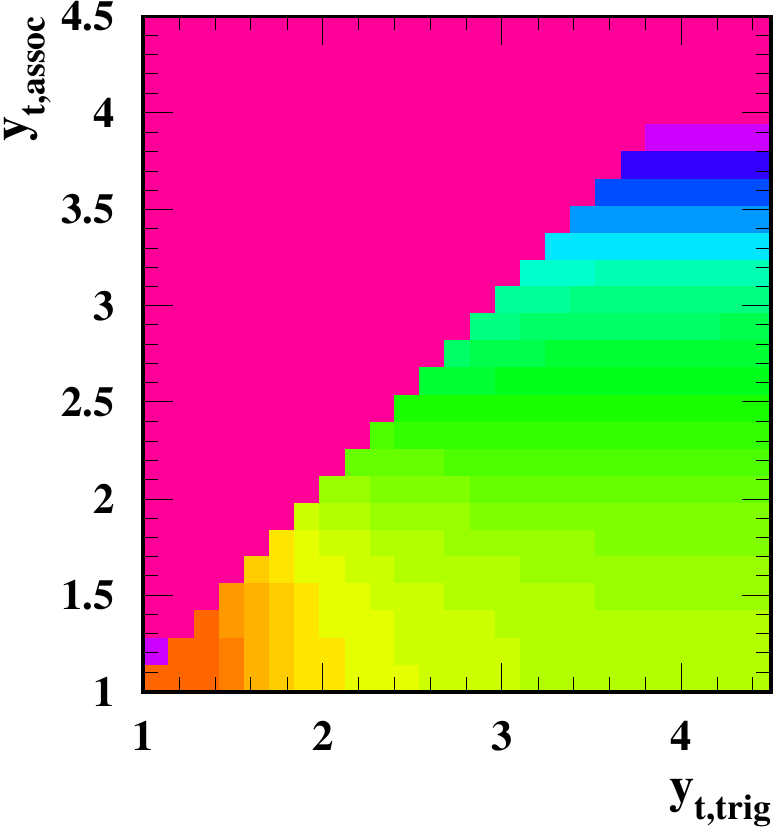}
\put(-82,90){\bf (b)}\\
\includegraphics[width=.47\columnwidth]{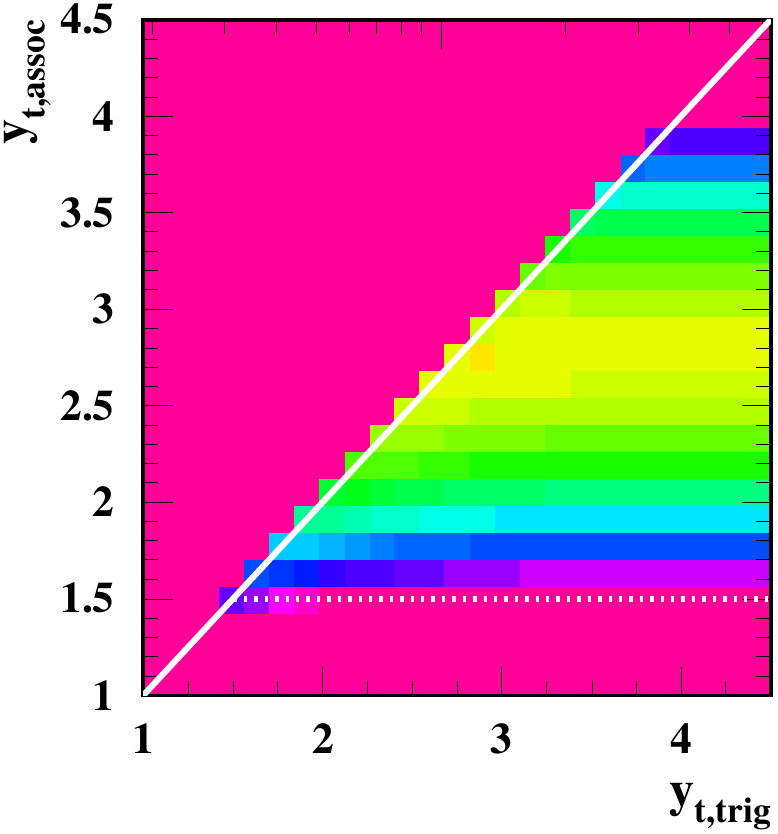}
\put(-82,90){\bf (c)}
\includegraphics[width=.48\columnwidth]{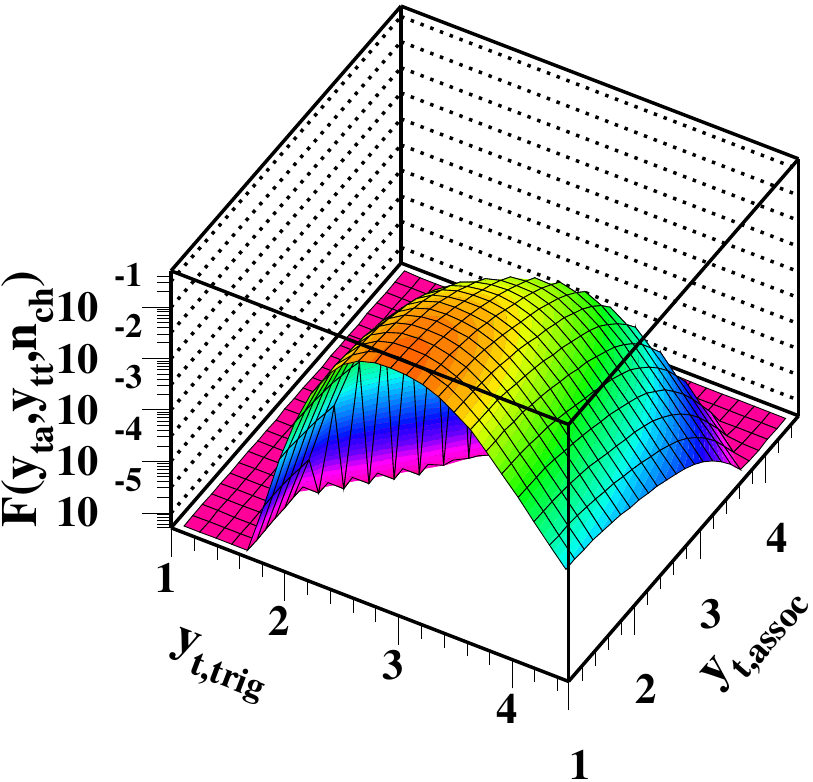}
\put(-85,90){\bf (d)}
\caption{
(a) Calculated TCM trigger spectrum $T(y_{t,trig})$ (solid curve) compared to the spectrum inferred from data from 200 GeV \pp\ collisions (solid points). Also shown are trigger spectra for soft $T_s$ (dotted) and hard $T_h$ (dashed) events.
(b) TCM for conditional-associated spectra $A(y_{t,assoc}|y_{t,trig})$.
(c)  TCM for conditional-associated hard component $A_{hh}(y_{t,assoc}|y_{t,trig})$.
(d) TCM for joint 2D TA correlations  $F(y_{t,assoc},y_{t,trig})$.
}
\label{tatcm1}       
\end{figure}

Figure~\ref{tatcm1} (a) shows a measured trigger spectrum (points) for event class $n=5$ compared to the corresponding TCM prediction (solid curve). The agreement is good. Other curves show soft $T_s$ and hard $T_h$ components of the trigger spectrum times corresponding event-type probabilities.
Figure~\ref{tatcm1} (b) shows the conditional associated component $A(y_{ta}|y_{tt})$ (for soft or hard events) obtained by sampling from the SP spectrum soft component with the condition $y_{ta} \leq y_{tt}$. For each $y_{tt}$ condition the associated-hadron  spectrum is normalized to $n_{ch} - 1$. Panel (c) shows the associated hard component (for hard events only)  $A_{hh}(y_{ta}|y_{tt})$ obtained from the SP spectrum hard component by following the same procedure.
Figure~\ref{tatcm1} (d) shows the combination of associated soft and hard components multiplied by the trigger spectrum to obtain the full TA joint distribution $F(y_{ta},y_{tt})$ TCM.

\begin{figure}
\centering
\includegraphics[width=.47\columnwidth]{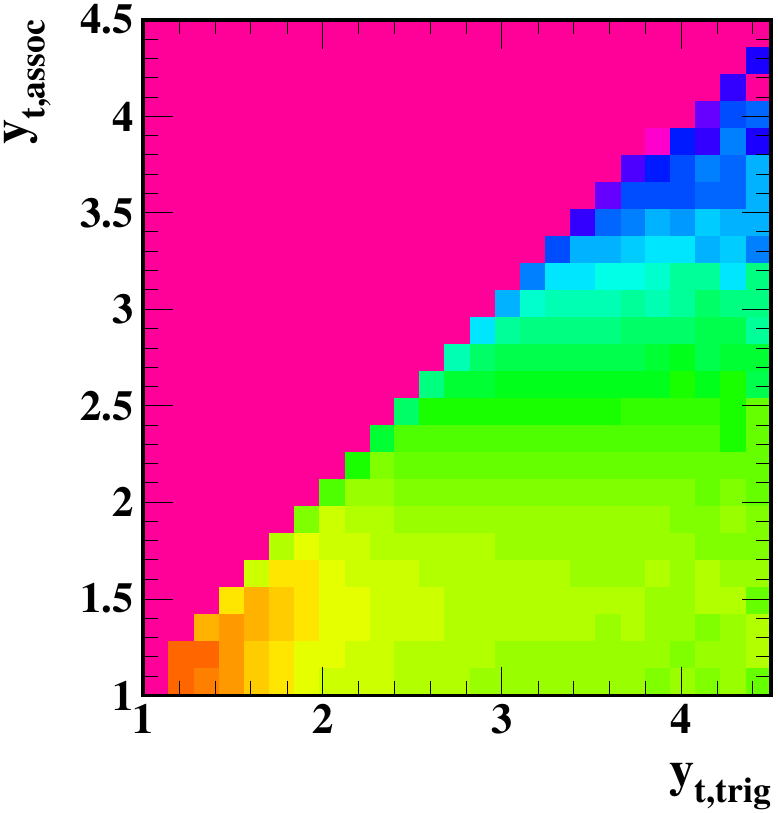}
\put(-82,92){\bf (a)}
\includegraphics[width=.47\columnwidth]{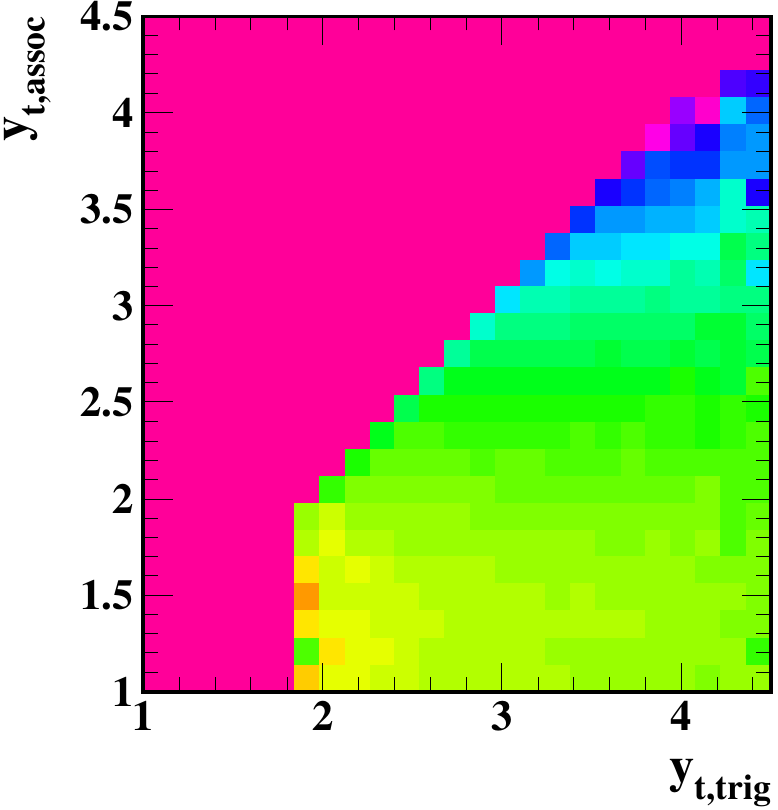}
\put(-82,92){\bf (b)}\\
\includegraphics[width=.99\columnwidth]{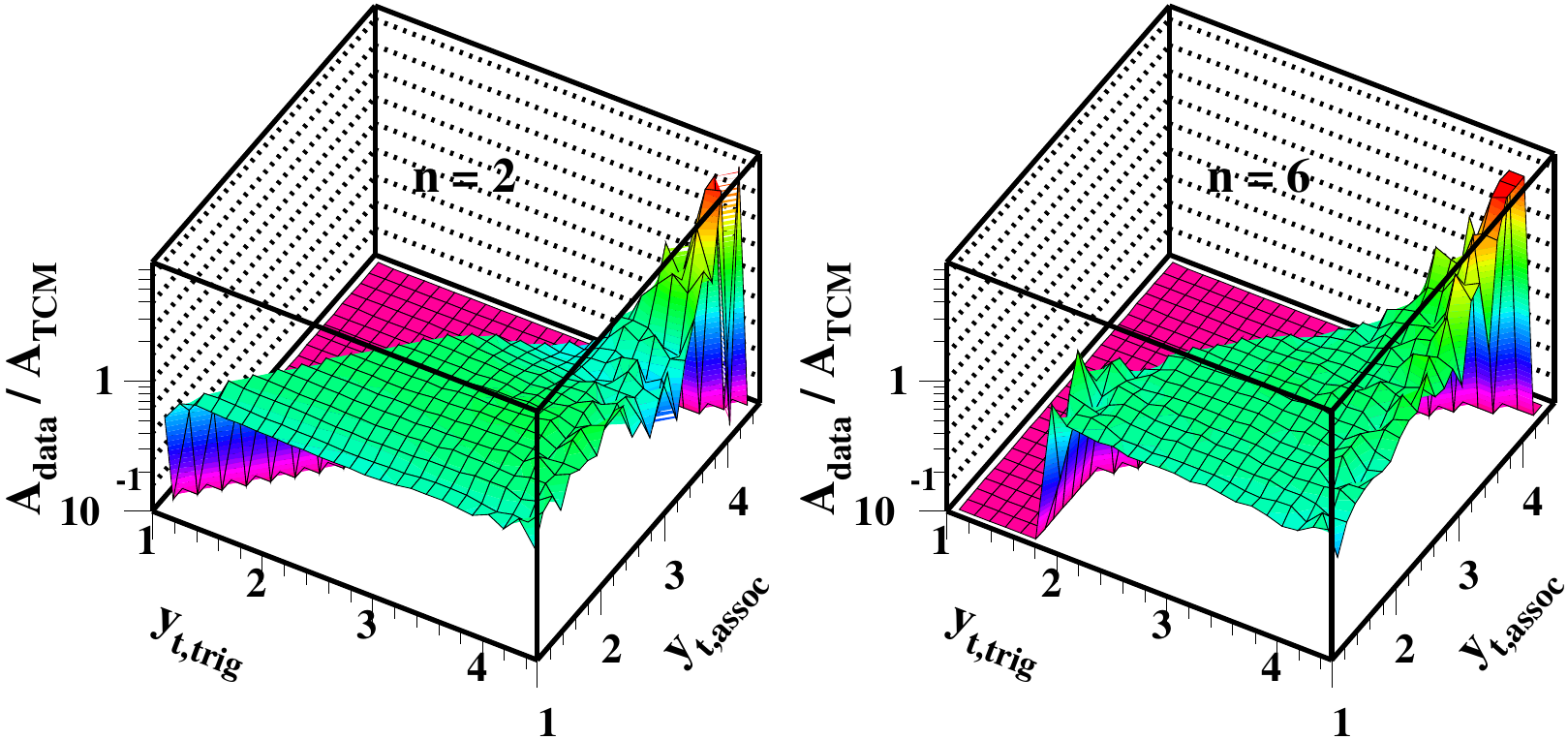}
\put(-85,92){\bf (d)}
\put(-202,92){\bf (c)}
\caption{
(a), (b) Conditional-associated TA correlations $A(y_{t,assoc}|y_{t,trig})$ for $n=2$, 6 multiplicity classes of 200 GeV \pp\ collisions.
(c), (d) Ratio data/TCM for $A(y_{t,assoc}|y_{t,trig})$ and for $n=2$, 6 multiplicity classes of 200 GeV \pp\ collisions.
}
\label{tatcm2}       
\end{figure}

Figure~\ref{tatcm2} (a,b) shows associated conditional distributions $A(y_{ta}|y_{tt})$ for two multiplicity classes obtained by dividing the measured joint distribution $F(y_{ta},y_{tt})$ by the measured trigger spectrum $T(y_{tt})$. The general similarity with Fig.~\ref{tatcm1} (b) is apparent. To further demonstrate the correspondence with the TA TCM panels (c,d) show data/model ratios. For $y_{ta} < 2.5$ the agreement is good (ratio near unity). The deviations from unity above that point are expected and reveal the jet-related correlation structure that is the object of this study.

%%%%%%%%%%%%%%%%%%
\section{Associated-conditional hard component}  \label{hardcomp}

The TA associated-conditional hard component of hard events can be extracted from measured TA correlations by subtracting TCM soft-component models in the form
\bea \label{hardcomp}
P_h R_hA_{hh}(y_{ta}|y_{tt},n_{ch}) \hspace{-.07in} &=& \hspace{-.07in} A - P_s R_s A_s -  P_h R_h A_{hs},
\eea
where the trigger ratios are $R_x = T_x / T$, and $P_s$ and $P_h$ are the estimated soft and hard event-type probabilities.

\begin{figure}
\centering
\includegraphics[width=.47\columnwidth]{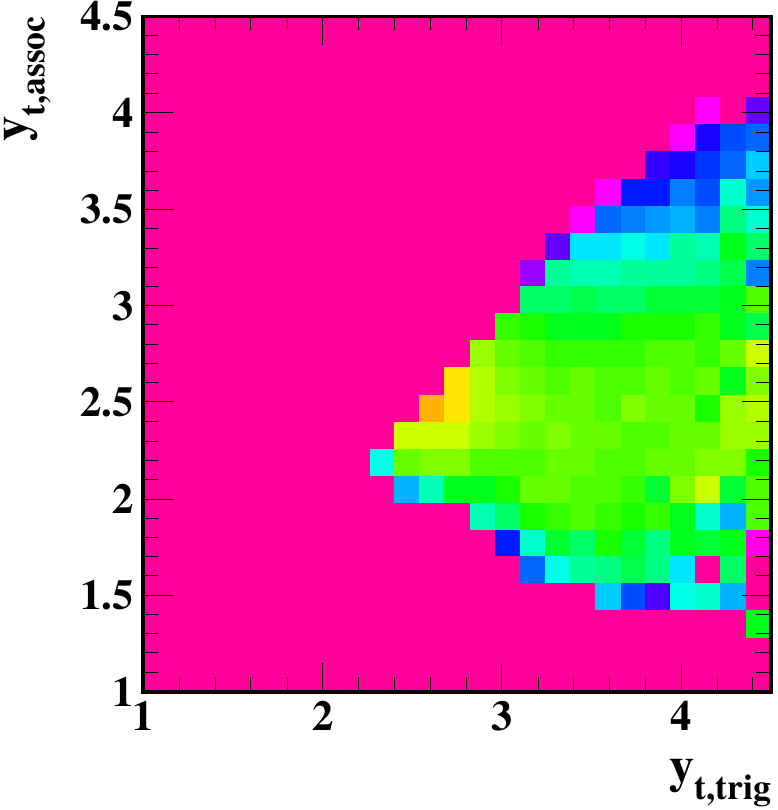}
\put(-82,92){\bf (a)}
\includegraphics[width=.47\columnwidth]{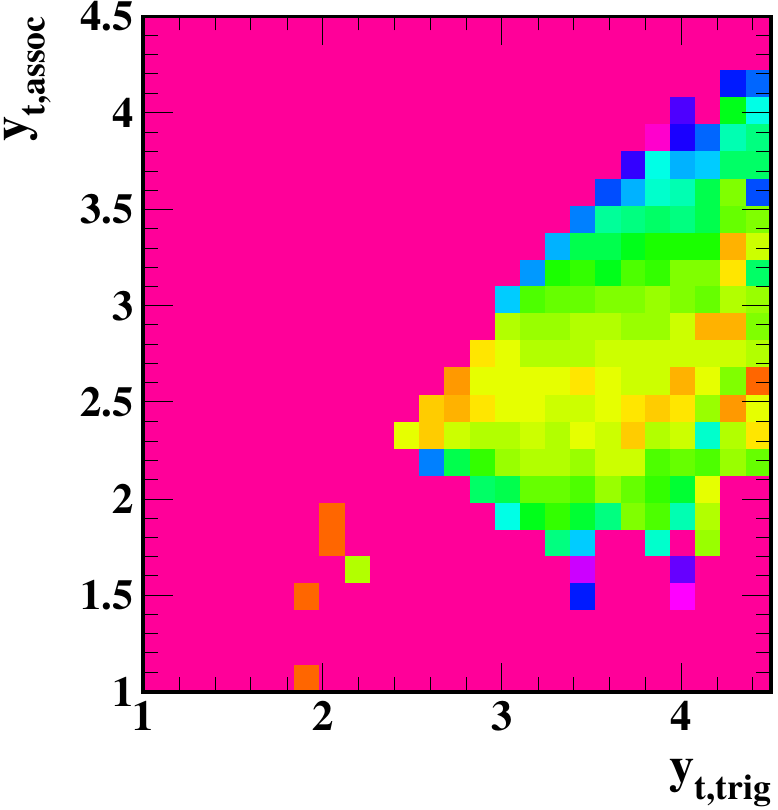}
\put(-82,92){\bf (b)}//
\includegraphics[width=.47\columnwidth]{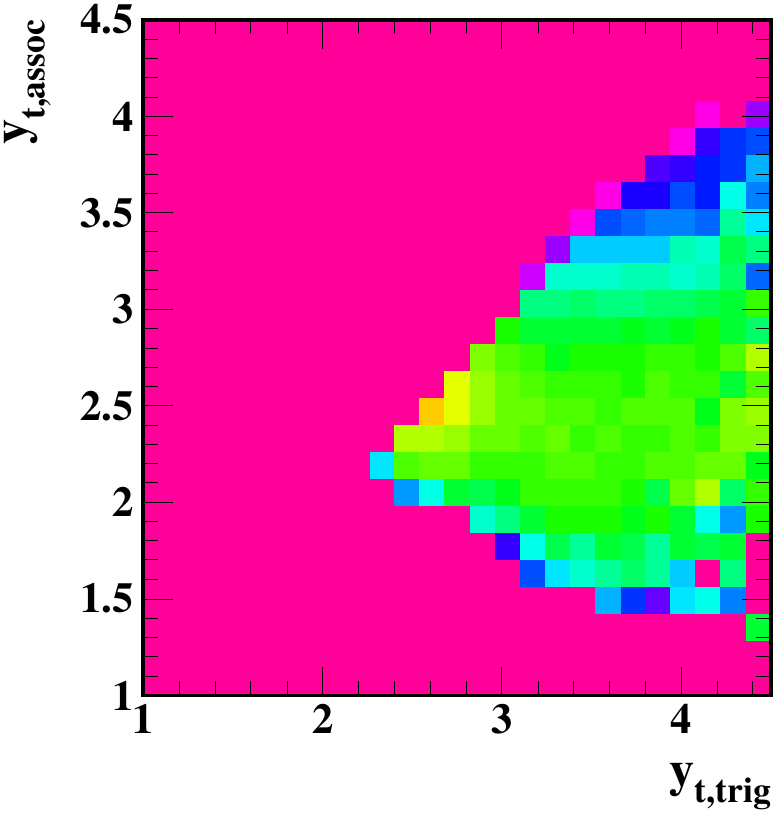}
\put(-82,92){\bf (c)}
\includegraphics[width=.47\columnwidth]{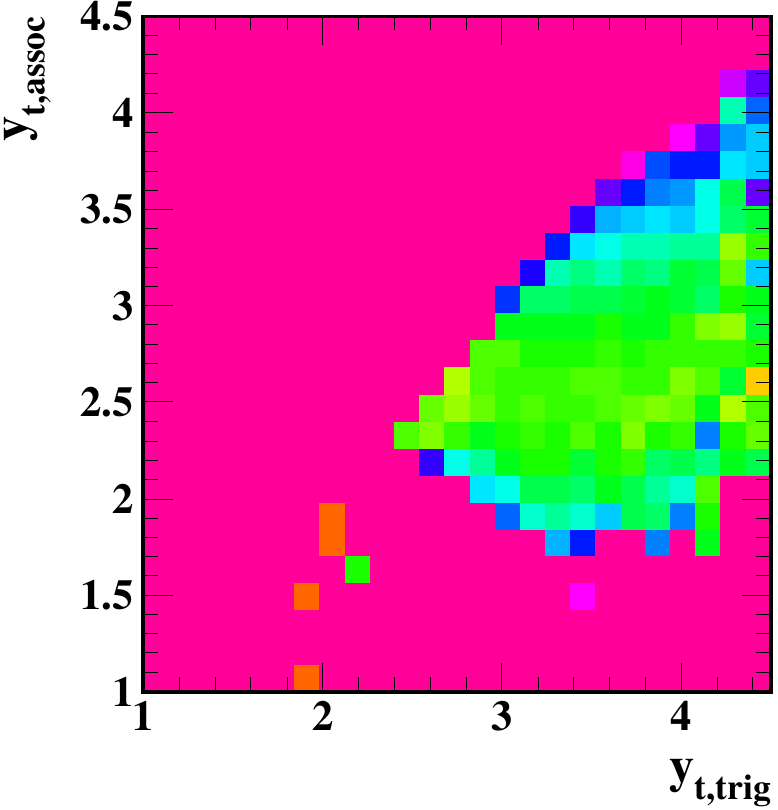}
\put(-82,92){\bf (d)}
\caption{
(a), (b) {\em Per-event} hard components of $A_{hh}(y_{t,assoc}|y_{t,trig})$ for multiplicity classes $n = 2$, 5 respectively of 200 GeV \pp\ collisions.
(c), (d)  {\em Per-dijet} hard components of $A_{hh}(y_{t,assoc}|y_{t,trig})$ for multiplicity classes $n = 2$, 5 respectively of 200 GeV \pp\ collisions based on trends in Fig.~\ref{ppspec} (d).
}
\label{tahard}       
\end{figure}

Figure~\ref{tahard} (a,b) shows associated conditional hard components $A_{hh}(y_{ta}|y_{tt})$ {\em from hard events} for two multiplicity classes. The amplitude increasing with event multiplicity confirms that the number of dijets increases well above 1, as in Fig.~\ref{ppspec} (d). Rescaling the amplitudes by the calculated dijet number from that panel we obtain panels (c,d): the jet-related correlations {\em per dijet} averaged over dijets does not change substantially over an \nch\ interval where the dijet number per event increases by a factor 2.

%%%%%%%%%%%%%%%%%%
\section{TA partition for fragmentation functions}  \label{ffs}

To establish a direct comparison between the system of fragmentation functions conditional on MB dijet spectrum and 2D TA correlations the FF system must be partitioned into trigger and associated components. The method follows that adopted to construct the TA TCM.

\begin{figure}
\centering
\includegraphics[width=.47\columnwidth]{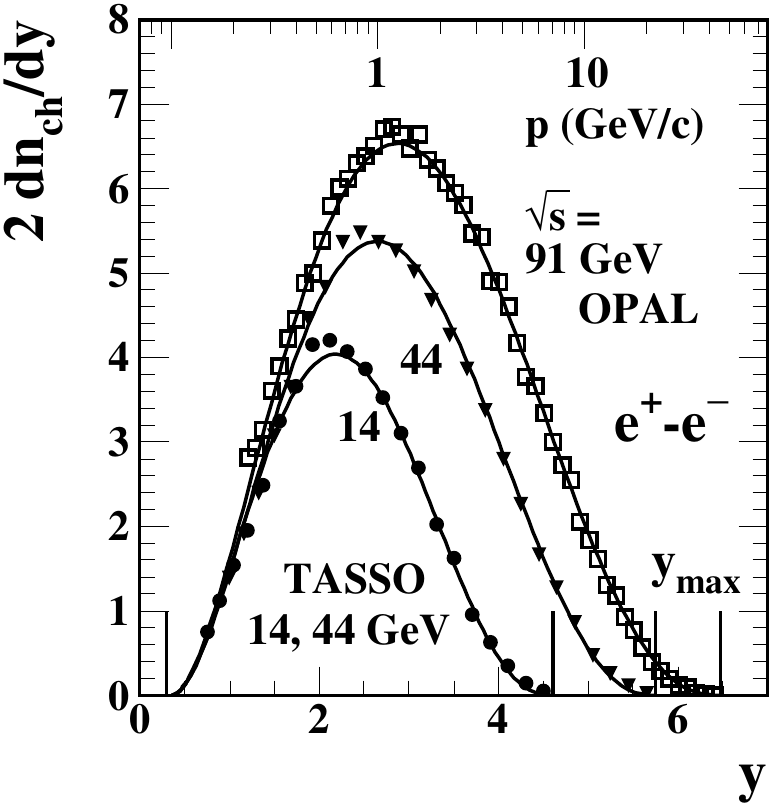}
\put(-82,92){\bf (a)}
\includegraphics[width=.47\columnwidth]{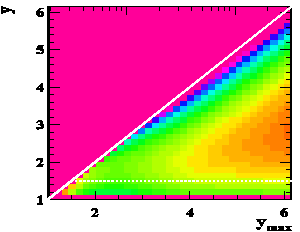}
\put(-82,92){\bf (b)}\\
\includegraphics[width=.47\columnwidth]{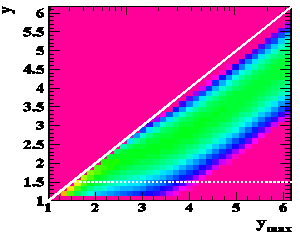}
\put(-82,92){\bf (c)}
\includegraphics[width=.47\columnwidth]{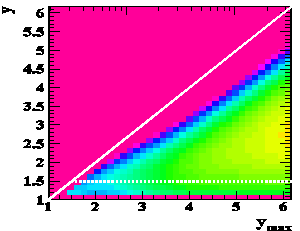}
\put(-82,92){\bf (d)}
\caption{
(a) Fragmentation functions (FFs) from TASSO~\cite{tasso} and OPAL~\cite{opal} plotted on rapidity $y=\ln[(E+p)/m_\pi]$ illustrating self-similar energy scaling in that format.
(b) Parametrization of \ee\ light-flavor quark FFs from 3 to 30 GeV based on the beta distribution (solid curves in previous panel)~\cite{eeprd}.
(c) Trigger-fragment component of  FF ensemble in panel (b).
(d) Associated-fragment component of  FF ensemble in panel (b).
}
\label{ffs}       
\end{figure}

Figure~\ref{ffs} (a) shows \ee\ FFs for three dijet energies plotted on fragment total rapidity $y = \ln[(p + E)/m_\pi]$. The data evolution with parton/jet energy or rapidity $y_{max} = \ln(2E_{jet}/m_\pi)$ is self-similar, the invariant shape described by a beta distribution. That regularity leads to a simple parametrization of FFs that is accurate over a large energy range (solid curves through the data). The complete FF ensemble for \ee\ collisions is shown as a surface in Figure~\ref{ffs} (b) extending up to 30 GeV jets (all RHIC jets at 200 GeV). The FF ensemble is separated into trigger and associated components by defining void probability $G(y_{trig}) = \exp[-n_\Sigma(y_{trig})]$ just as for the TA TCM. 
 \bea \label{stdu}
S_t(y_{trig}|y_{max}) \hspace{-.07in} &=&\hspace{-.07in} G_t(y_{trig}|y_{max}) \epsilon(\Delta \eta) D(y_{trig}|y_{max}),
 \eea
where factor $ \epsilon(\Delta \eta)$ is required also in determining the void probability because $G$ depends on the number of fragments actually accepted.
The trigger $S_t(y_{trig}|y_{max})$ and associated $D_a(y_{assoc}|y_{max})$ components of the full FFs $D(y|y_{max})$ are shown for gluon jets from \ee\ collisions in Figure~\ref{ffs} (c,d). Quark-jet FFs are treated similarly. A similar partition of \ppbar\ FFs is also incorporated in this analysis.

%%%%%%%%%%%%%%%%%%
\section{FF convolutions and Bayes' theorem}  \label{bayes}

To provide a direct comparison between FF systematics and jet-related TA correlations requires transforming from trigger and associated fragments both conditional on parton energy to associated fragments conditional on trigger fragments by integrating over the parton degree of freedom. The transformation makes use of Bayes' theorem.

\begin{figure}
\centering
\includegraphics[width=.45\columnwidth]{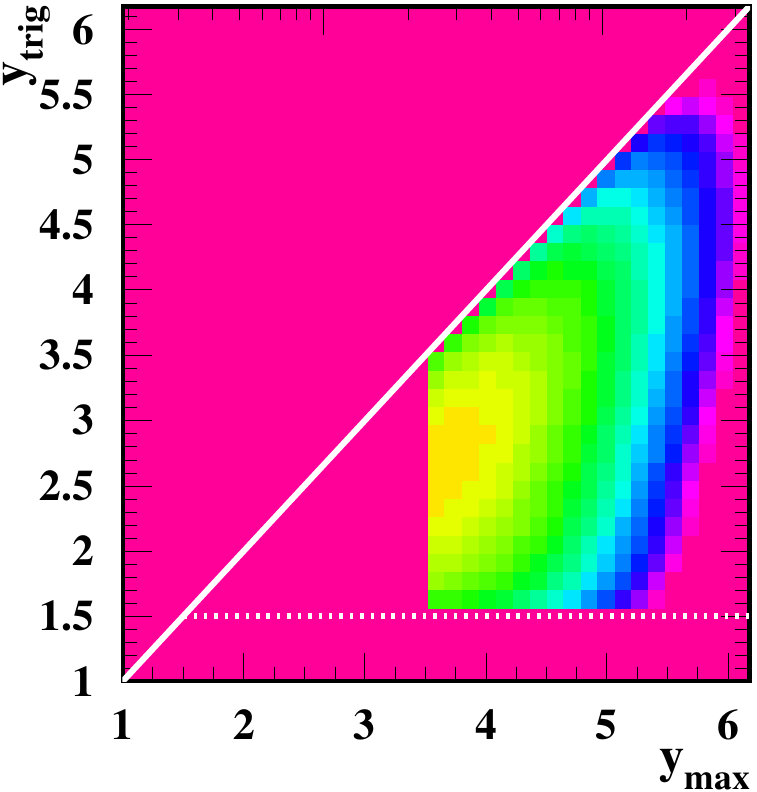}
\put(-82,52){\bf (a)}
\includegraphics[width=.48\columnwidth]{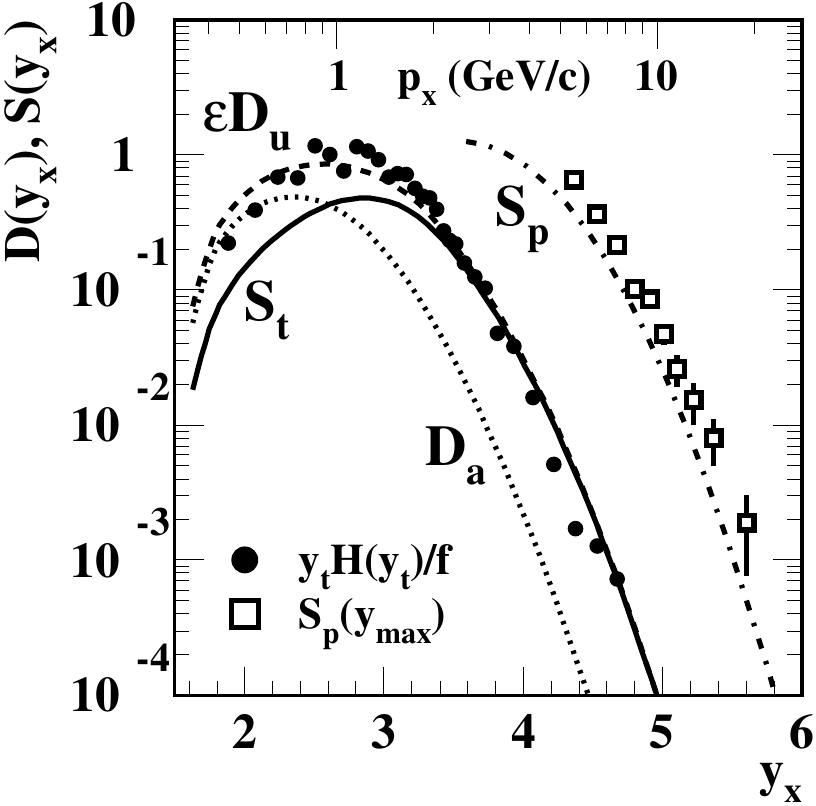}
\put(-82,52){\bf (b)}\\
\includegraphics[width=.46\columnwidth]{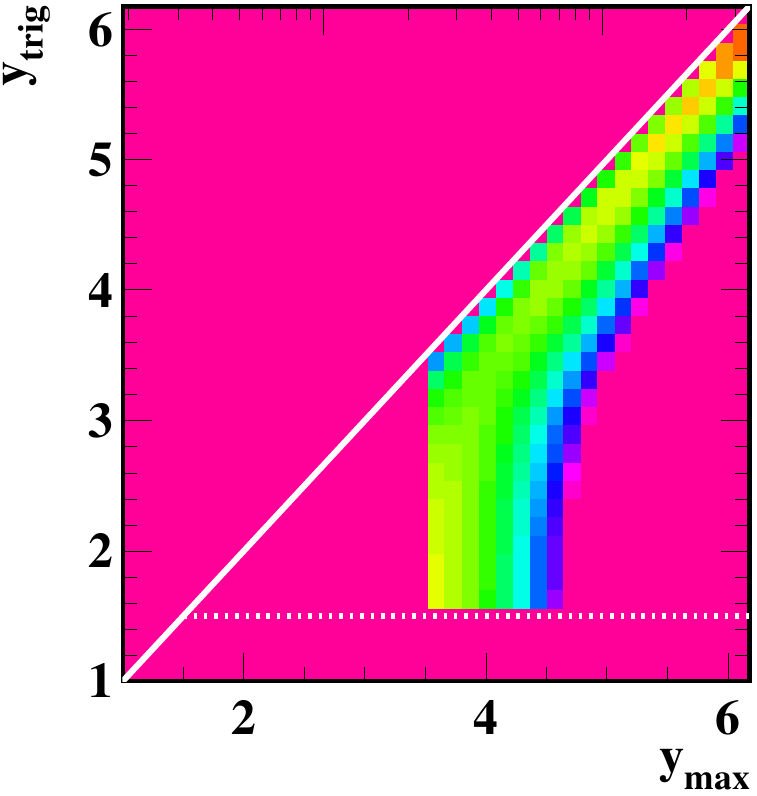}
\put(-82,52){\bf (c)}
\includegraphics[width=.49\columnwidth]{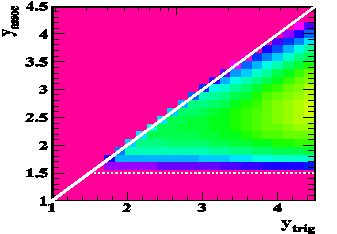}
\put(-82,52){\bf (d)}
\caption{\label{convolute}  
(a) Joint trigger-parton distribution  $F_{tp}(y_{trig},y_{max})$ obtained by combining the FF trigger component in Fig.~\ref{ffs} (c) with measured MB jet (scattered parton) spectrum $S_p$ in panel (b).
(b) Projection $S_t(y_{trig})$ of the joint distribution in panel (a) onto $y_{trig}$ (solid curve) compared to the \pp\ spectrum hard component (per dijet) in the form $y_t H(y_t) / f$ (solid points).
(c) Parton spectrum ensemble $S_p(y_{max}|y_{trig})$ conditional on trigger rapidity, derived from Eq.~(\ref{eq66}).
(d) The sought-after FF associated-conditional distribution $D_a(y_{assoc}|y_{trig})$ defined in Eq.~(\ref{xxx}) that can be compared  with the corresponding TA hard component $A^*_{hh}(y_{t,assoc}|y_{t,trig})$ from triggered dijets. This example is for quark FFs from \ppbar\ collisions.
}
     
\end{figure}

We first obtain the  trigger-fragment spectrum averaged over the MB jet spectrum $S_p(y_{max})$.
The FF trigger-parton joint distribution is $F_{tp}(y_{trig},y_{max}) = S_p(y_{max})  S_t(y_{trig}|y_{max})$ shown in Fig.~\ref{convolute} (a). The marginal projection onto $y_{trig}$ is the trigger-fragment spectrum 
 \bea \label{sthat}
S_t(y_{trig}) &=& \int_{y_{trig}}^\infty dy_{max}    S_p(y_{max})  S_t(y_{trig}|y_{max})
\\ \nonumber
&\equiv& G_t(y_{trig})\epsilon(\Delta \eta) D_u(y_{trig}),
 \eea
defining marginal void probability $G_t(y_{trig})$ as a spectrum-weighted average of the $G_t(y_{trig}|y_{max})$. 
Figure~\ref{convolute} (b) shows $S_t(y_{trig})$ (solid curve) as the marginal projection of $F_{tp}(y_{trig},y_{max})$ onto $y_{trig}$ and the parton spectrum $S_p(y_{max})$ as the marginal projection onto $y_{max}$ (dash-dotted curve) based on measured jet spectra (e.g.\ open squares).  Similarly, $F_{ap}(y_{assoc},y_{max}) =S_p(y_{max}) D_a(y_{assoc}|y_{max})$ has marginal projection $D_a(y_{assoc})$ integrating to $\epsilon 2 \bar n_{ch,j} - 1$.

To proceed we use a relation represented schematically by
\bea \label{foldd}
F_{at}(a,t) &\approx& \int dy_{max} F_{ap}(a,p) F_{tp}(t,p) / F_p(p) \\ \nonumber
&\rightarrow& \int dy_{max}  F_{ap}(a,p) F_{pt}(p,t) / S_p(p) 
\eea
 to remove by integration the parton degree of freedom and obtain FF TA correlations. Converting all the joint distributions to the form $A(x,y)=B(x|y)C(y)$ via the chain rule and canceling common marginal distributions we obtain
\bea \label{xxx}
 \hspace{-.22in}
D_a(y_{assoc}|y_{trig}) \hspace{-.1in} &\approx&  \hspace{-.15in} \int \hspace{-.05in} dy_{max}  D_{a}(y_{assoc}|y_{max})  S_{p}(y_{max}|y_{trig}).~~~~
\eea
We require the missing conditional $S_{p}(y_{max}|y_{trig})$ that may be obtained by an application of Bayes' theorem
\bea \label{eq66}
S_{p}(y_{max}|y_{trig})  \approx \frac{S_t(y_{trig}|y_{max}) S_p(y_{max})}{ S_t(y_{trig})}.
\eea

Figure~\ref{convolute} (c) shows the conditional distribution $S_{p}(y_{max}|y_{trig})$.
Note the difference between $S_t(y_{trig}|y_{max}) $ in Fig.~\ref{ffs} (c) and  $S_{p}(y_{max}|y_{trig})$ in Fig.~\ref{convolute} (c).
Figure~\ref{convolute}  (d) shows the sought-after $D_a(y_{assoc}|y_{trig})$ (in this case for \ppbar\ quark FFs) that can be compared with TA hard components from \pp\ collisions. 
Those trigger-associated distributions derived from FFs for \ee\ and \ppbar\ collisions and in each case for gluon jets and light-flavor quark jets (four combinations) can be compared directly with TA hard components $A^*_{hh}(y_{t,assoc}|y_{t,trig})$ for triggered dijets.

\begin{figure*}[t]
\centering
\includegraphics[width=.649\columnwidth]{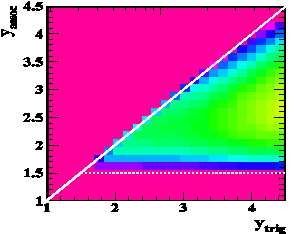}
\put(-115,120){\bf (a)}
\includegraphics[width=1.2\columnwidth]{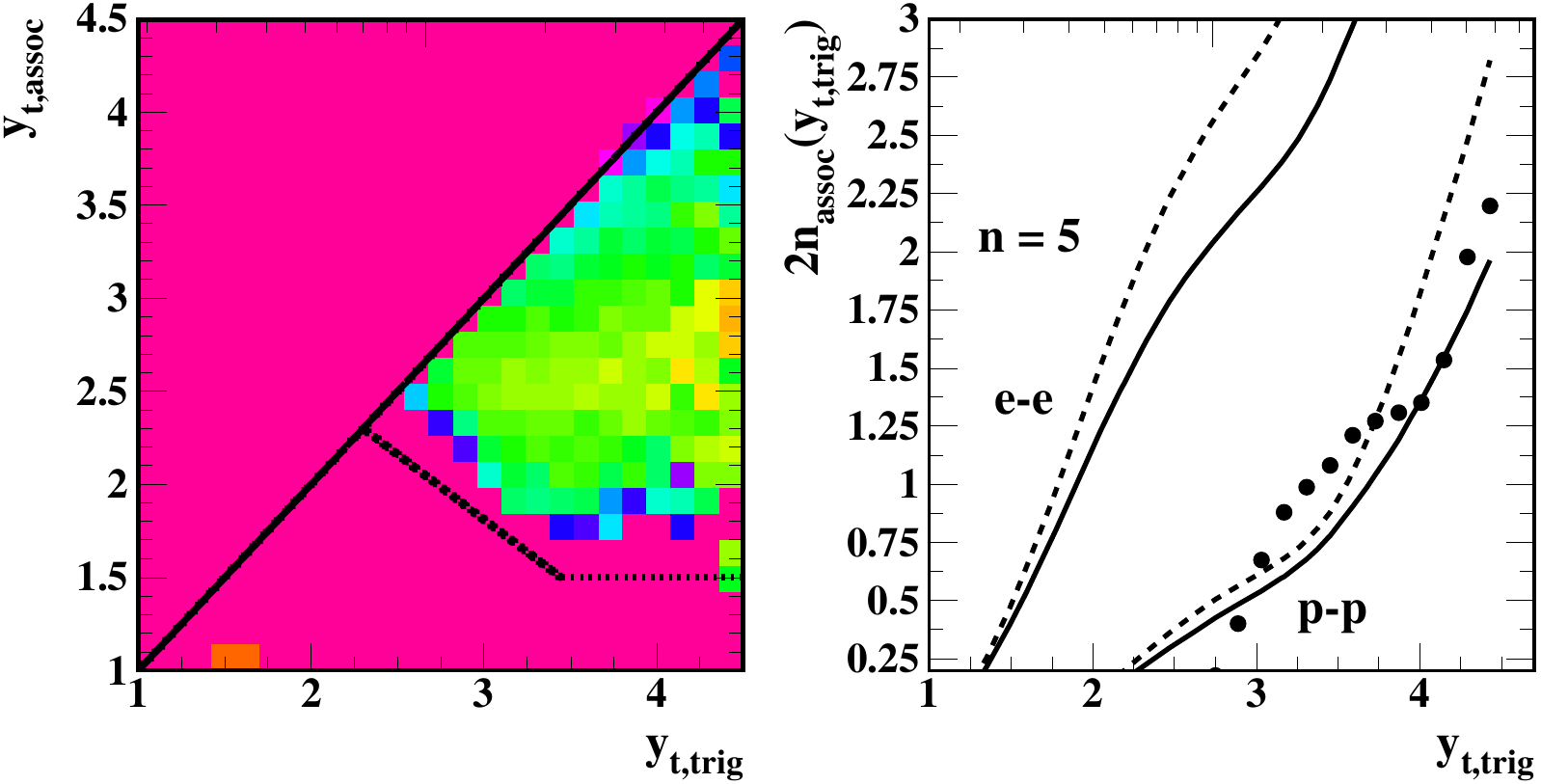}
\put(-100,120){\bf (c)}
\put(-240,120){\bf (b)}
\caption{
(a) Repeat of Fig.~\ref{convolute} (d) for comparison with \pp\ data.
(b) TA conditional-associated hard component $A^*_{hh}(y_{t,assoc}|y_{t,trig})$ corresponding to triggered (primary) dijets. The z axis (log scale) is the same as in panel (a).
(c) Projection of FF calculations onto $y_{trig}$ for LF quark (solid) and gluon (dashed) jets and for \ppbar\ (lower) and \ee\ (upper) collisions compared to data from panel (b) (points).
}
\label{hardcomp1}       
\end{figure*}

%%%%%%%%%%%%%%%%%%
\section{QCD predictions vs TA data}  \label{qcd}

We are now in a position to compare TA hard-component data from 200 GeV \pp\ collisions with QCD predictions derived from reconstructed jets. I am interested in comparisons of correlation structure, amplitudes (z axis), possible kinematic boundaries and 1D projections onto $y_{t,trig}$ and $y_{t,assoc}$. Note that whereas the TA associated {\em total} is constrained to $n_{ch} - 1$ when projected onto $y_{t,trig}$ (e.g.\ Fig.~\ref{tatcm2}) the TA hard component is not. Each column sums to the physical hadron fragment number modulo a small bias.

Figure~\ref{hardcomp1} (a) repeats the fourth panel of Fig.~\ref{convolute} showing the TA associated hard component $D_{a}(y_{assoc},y_{trig})$ derived from \ppbar\ dijet FFs as in the previous section. Panel (b) shows $A^*_{hh}(y_{t,assoc},y_{t,trig})$, the part of $A_{hh}$ corresponding to a hard trigger correlated with the dijet to which it belongs (as opposed to uncorrelated contributions from soft triggers and secondary dijets that have been subtracted from $A_{hh}$ according to the TCM). Those data should be directly comparable with the TA constructions from single-jet FFs. The z-axis scales are the same. The amplitudes and general structure correspond well over large momentum and energy intervals. There is some disagreement for the smallest hadron rapidities which provides information on the real kinematic limits to jet formation.

Figure~\ref{hardcomp1} (c) shows projections of data from the first panel (lower solid curve) and second panel (points) onto $y_{t,trig}$ ($y_{trig}$ for the first panel). The other curves correspond to calculations for gluon jets from \ee\ and \ppbar\ collisions (dashed) and quark jets from \ee\ collisions (upper solid curve). The comparison indicates that FFs from \ppbar\ collisions are strongly  favored by the TA correlation data.

\begin{figure*}
\centering
\includegraphics[width=.628\columnwidth]{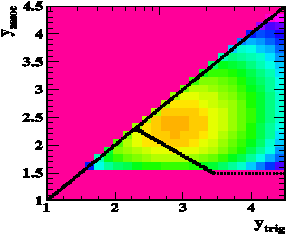}
\put(-110,120){\bf (a)}
\includegraphics[width=1.2\columnwidth]{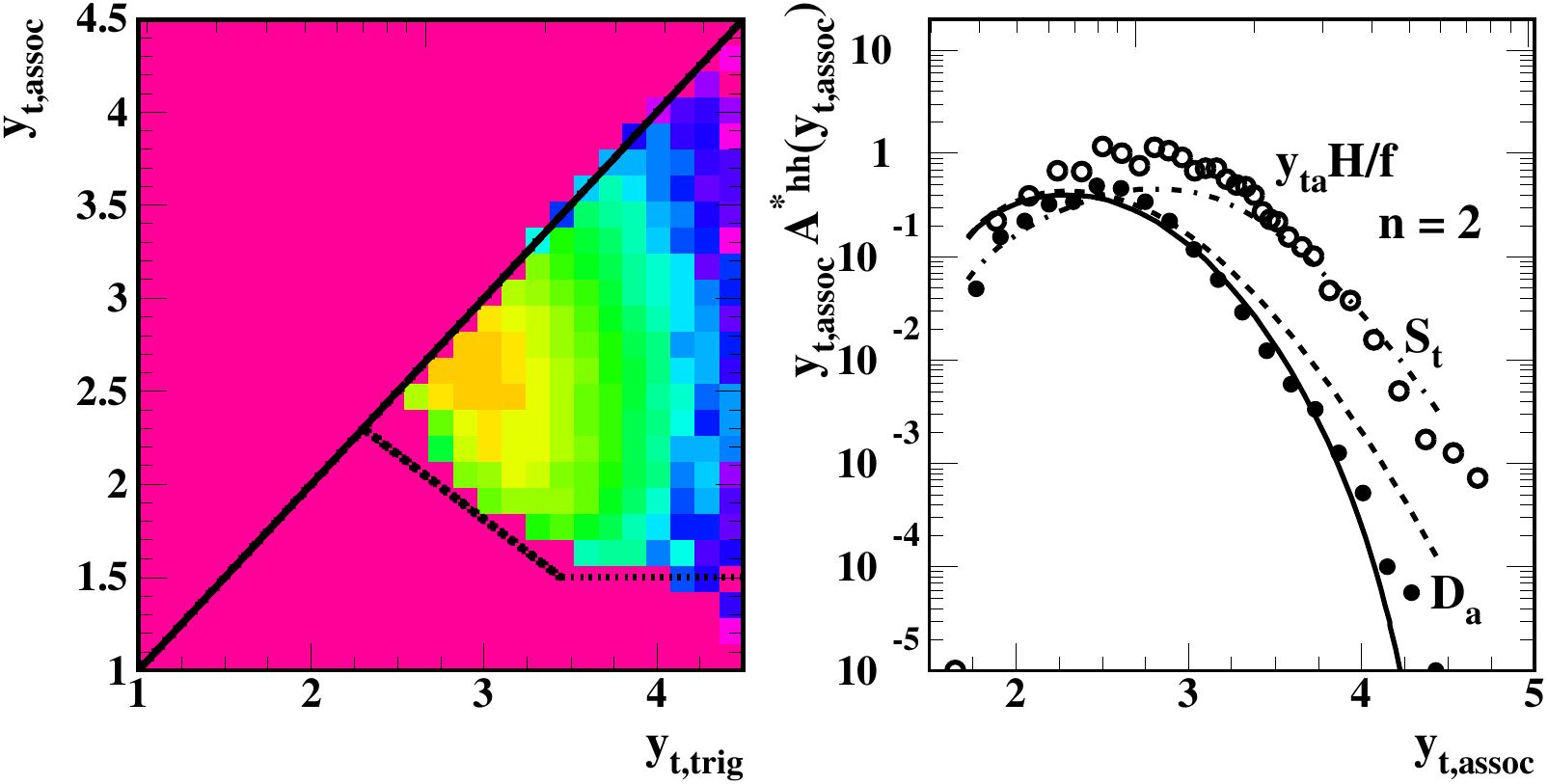}
\put(-100,120){\bf (c)}
\put(-240,120){\bf (b)}
\caption{
(a) Predicted joint TA distribution $F_{at}(y_{assoc},y_{trig}) = S_t(y_{trig}) D_a(y_{assoc}|y_{trig})$  from FFs and a MB dijet spectrum.
(b) Measured TA joint distribution $F^*_{at}(y_{t,trig},y_{t,assoc}) = T_h(y_{t,trig}) A^*_{hh}(y_{t,assoc}|y_{t,trig})$ corresponding to triggered (primary) dijets. The z axis (log scale) is the same as in panel (a).
(c)  Projection of FF calculations onto $y_{assoc}$ for LF quark jets from  \ppbar\  collisions (solid curve)  compared to $D_a(y_{t,assoc})$, the projection of data from panel (b) onto $y_{t,assoc}$ (solid points) and the \pp\ MB spectrum hard component (open points).
}
\label{hardcomp2}       
\end{figure*}

Figure~\ref{hardcomp2} (a) shows the predicted joint trigger-fragment distribution $F_{at}(y_{assoc},y_{trig}) = S_t(y_{trig})D_{a}(y_{assoc},y_{trig})$ based on FFs from \ppbar\ collisions. The most-probable trigger fragment appears near 1 GeV/c, and the most probable associated fragment near 0.75 GeV/c. Panel (b) shows the corresponding data in the form $F_{hh}((y_{assoc},y_{trig})) = T_h(y_{t,trig})A_{hh}(y_{t,assoc},y_{t,trig})$. Again the z-axis scales are the same and there is general agreement in amplitude and structure. The disagreements for smaller hadron momenta reveal real kinematic bounds on jet formation that are not reflected in the prediction that assumes QCD factorization. These distributions represent the absolute yield of MB jet-related hadron pairs and {\em based on measured jet properties} demonstrate that the great majority of jet fragments appears below 2 GeV/c, contradicting conventional assumptions about the final-state structure of high-energy nuclear collisions.

Figure~\ref{hardcomp2} (c) shows projections of the first panel (solid curve) and second panel (solid points) onto $y_{assoc}$ and $y_{t,assoc}$ respectively. The projection onto $y_{assoc}$ is $D_a(y_{assoc})$, the MB mean associated-fragment distribution mentioned at the end of Sec.~\ref{bayes}. Also shown is the trigger-fragment spectrum $S_t(y_{trig})$. The two together must sum to $D_u(y)$, the MB mean jet fragment distribution~\cite{fragevo,jetcorr} that is in turn comparable with the \pp\ spectrum hard component in the form $y_t H(y_t) / f(n_s)$ (open points) from Refs.~\cite{ppprd,fragevo}. The quantitative agreement of the predicted 2D structure and 1D projections with measured spectra and correlations reveals a self-consistent theory-data system in which MB dijets play a dominant role. 

Note that so-called ``dihadron correlation analysis'' on 1D azimuth based on trigger and associated \pt\ cuts~\cite{dihadron} typically acknowledges only a very small fraction of the total jet fragment population, for instance a rectangle at upper right in Fig.~\ref{hardcomp2} (b) bounded by $y_{t,assoc} \in [3.3,4]$ ($\approx$ [2,4] GeV/c) and $y_{t,trig} \in [4,4.5]$ ($\approx$ [4,6] GeV/c), including much less than 1\% of all MB jet fragments.

%%%%%%%%%%%%%%%%%%
\section{Implications for MPI}  \label{mpi}

Given the trigger-particle azimuth direction as the proxy for a dijet axis (with some probability) we can separate associated particles into several azimuth intervals, as in conventional underlying event (UE) studies. In such studies the triggered dijet is assumed to be confined to ``toward'' $|\phi - \phi_{trig}| \in [0,\pi/3]$ and ``away'' $|\phi - \phi_{trig}| \in [2\pi/3,\pi]$ intervals. The excluded interval(s) is referred to as the ``trans'' (TR) region and is assumed to excluded any contribution from the triggered dijet~\cite{rick,cdfue,cmsue,pptheory}. The TR region should then provide access to the UE, defined as what is not included in the triggered dijet including beam-beam remnants (BBR, projectile nucleon fragments). In particular, secondary dijets from multiple parton interactions (MPI)~\cite{mpi} uncorrelated with the trigger may contribute to the TR.

\begin{figure*}
\centering
\includegraphics[width=.6\columnwidth]{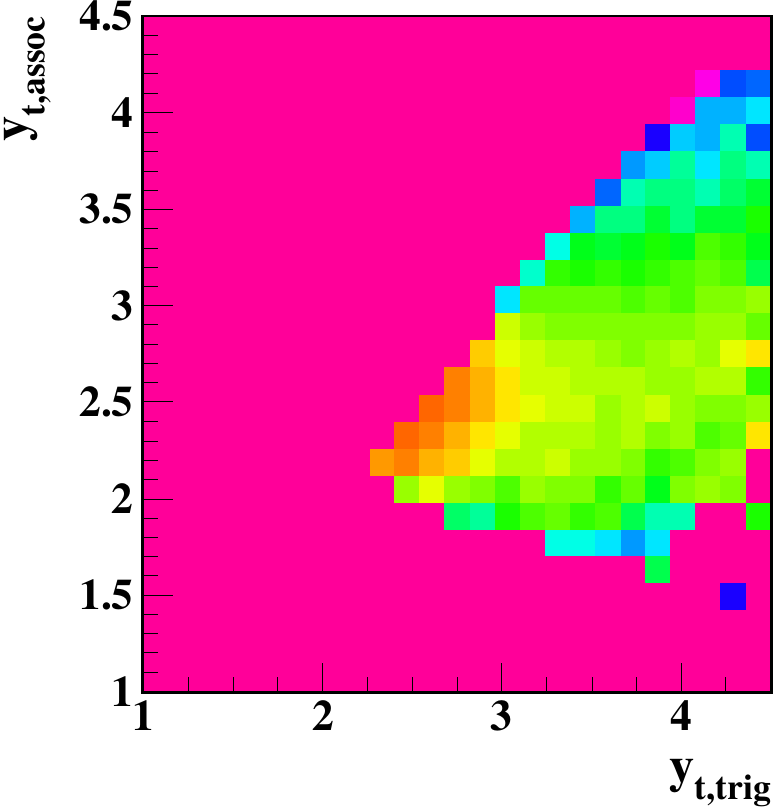}
\put(-100,120){\bf (a)}
\includegraphics[width=.6\columnwidth]{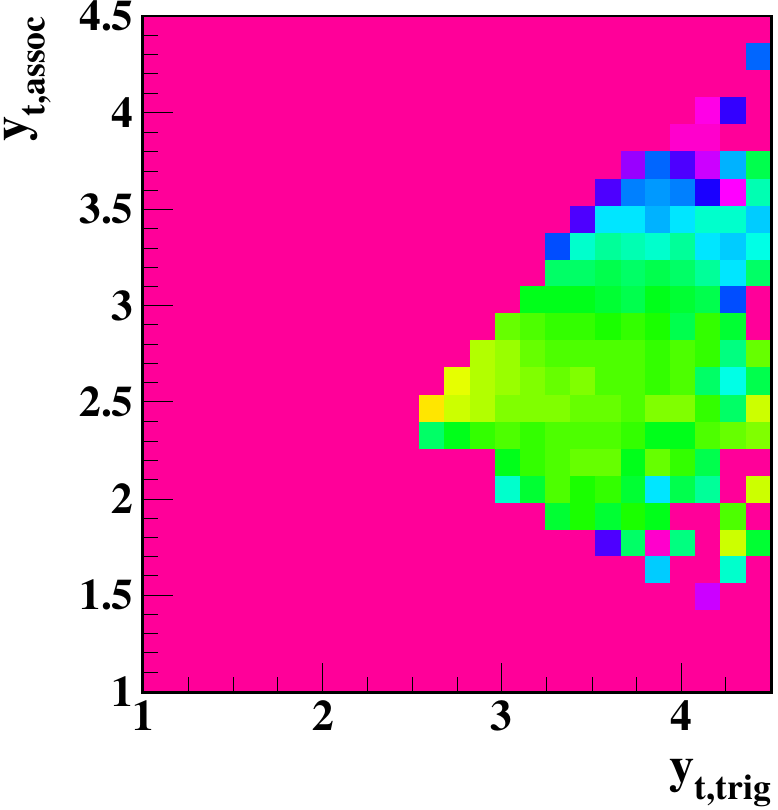}
\includegraphics[width=.6\columnwidth]{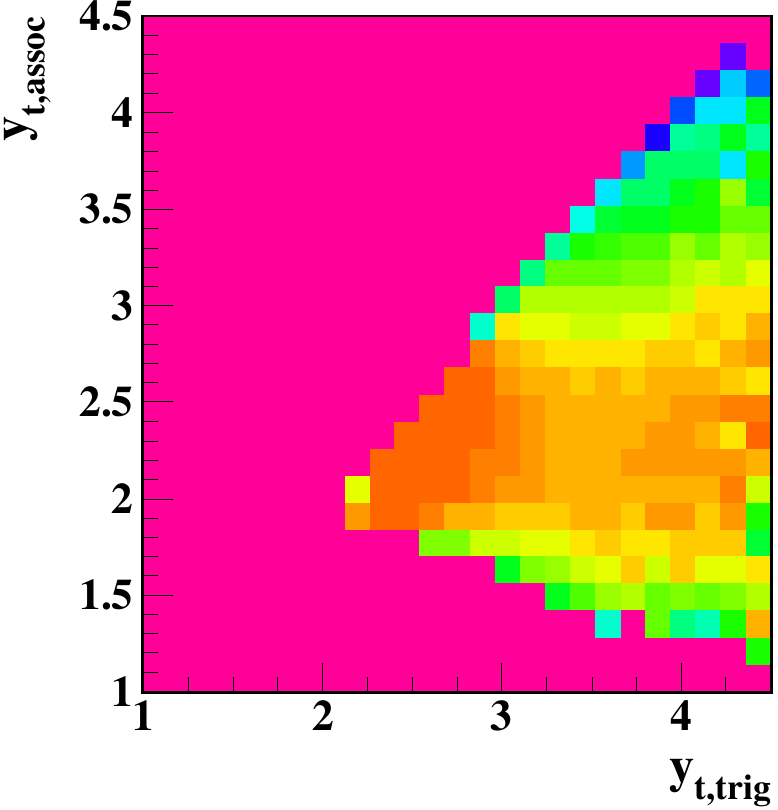}
\put(-100,120){\bf (c)}
\put(-240,120){\bf (b)}
\caption{
TA conditional hard components $A_{hh}(y_{t,assoc}|y_{t,trig})$ for low-multiplicity events (small probability of secondary dijets) and for three azimuth regions relative to the trigger direction --
(a) toward,
(b) trans,
(c) away --
showing a substantial contribution to the trans region from the triggered dijet.
}
\label{ue}       
\end{figure*}

Figure~\ref{ue} shows associated-conditional distributions from the three azimuth regions (a) toward, (b) trans and (c) away respectively. Figure~\ref{tahard} (a) is essentially the azimuth average over those three panels.  The small event multiplicities for this class insure that the probability of a second dijet is small (less than 10\%). The prominent jet  structure in toward (a) and away (c) intervals is expected. However, the substantial dijet contribution in the TR region (b) is not consistent with standard UE assumptions.

Based on MB correlation studies of \pp\ and \auau\ collisions the triggered dijet should actually contribute significantly to the TR region, as in Refs.~\cite{porter2,porter3,anomalous}. In the MB case the same-side and away-side dijet peaks on azimuth are sufficiently broad that they overlap by a considerable amount, thus contributing substantially to the TR region. That result is consistent with the picture of a dijet as radiation from a color dipole in which substantial radiation (ultimately hadrons) should emerge near the midplane of the dipole, corresponding to the TR region in a nuclear collision. If higher-energy jets are triggered the additional hadron fragments appear near the dijet axis, giving the impression that the SS and AS peaks are localized near 0 and $\pi$. However, the MB dijet base remains a common part of all dijets no matter what the jet energy. The presence of the MB dijet contribution to the TR has caused some confusion in interpretation of TR systematics, including inference of a substantial MPI contribution to low-multiplicity \pp\ collisions that actually arises from the triggered dijet. It is apparent from Fig.~\ref{ppspec} (d) that event multiplicity controls the dijet rate and thus the MPI rate, not a dijet trigger.

%%%%%%%%%%%%%%%%%%
\section{Discussion}  \label{disc}

Several lessons can be derived from this quantitative comparison of predicted TA correlations derived from measured FFs and jet spectra with measured TA correlations.

\subsection{The importance of low-energy jets}

In many analyses of the hadronic final state of high-energy nuclear collisions jet-related structures fully consistent with jet measurements and the general principles of QCD are interpreted instead as flow phenomena, based on analysis with data models motivated by strong flow assumptions and interpretations imposed from a flow context. Information in the data is typically suppressed in favor of preconceived notions, by projection to subspaces, by strong \pt\ cuts eliminating most hadrons and pairs, by defining ratios that obscure jet contributions and by ignoring collision centralities where jet structure is most obvious. We should instead establish as differentially and comprehensively as possible the real network of jet manifestations in yields, spectra and correlations from high-energy nuclear collisions and explicit phenomenology of event-wise-reconstructed jets and dijets in \pp, \aa, and \ee\ collisions. This study of TA correlations comparing QCD jet theory to data is an extension of a continuing program to establish a MB jet reference for all nuclear collisions. The consistent message from all such studies has been that dijets make a strong contribution to hadron production at lower hadron momentum, increasingly so in higher-multiplicity \pp\ and more-central \aa\ collisions. Descriptions of nuclear collisions that omit that large dijet component cannot be successful and should be excluded.

\subsection{How well does the TCM describe nuclear data?}

The TCM has been applied to yields, spectra, angular and $y_t \times y_t$ correlations and now to TA correlations. The TCM has previously been related quantitatively to pQCD calculations based on measured dijet production and fragment distributions, and quantitative agreement has been observed in all cases. Such comparisons rely on an accurate and convenient parametrization of quark and gluon FFs obtained separately from \ee\ and \ppbar\ collisions. They also rely on the recent development of a universal parametrization of jet spectra that describes spectra accurately for all \pp\ collision energies below 2 TeV. 

In the present study a theory prediction for the jet-related hard component of measured TA correlations is established for the first time. The description relies on a more elaborate algebraic system than previous theory comparisons with spectrum structure.  From FFs as conditional distributions on fragment rapidity conditional on parton rapidity we obtain analogous distributions of associated hadrons conditional on trigger hadrons that can be compared with TA hard-component data.

An error in any step of a long sequence could have resulted in falsification. The probability chain rule could have been inappropriate, as could application of Bayes' theorem. And the partition of FFs into trigger and associated components might have been mistaken. But the final comparison between predictions and data shows agreement at the 10\% level with no adjustable parameters. It seems that various assumptions were basically correct and the use of measured jet properties was in fact appropriate.

Comparing theory with 2D data in Figs.~\ref{hardcomp1} and \ref{hardcomp2} we find that as for 1D spectrum comparisons we again encounter quantitative agreement except for the lowest hadron momenta where we discover the real kinematic bounds on jet formation that could not be anticipated by the prediction simply because there has been no previous information on that nonperturbative aspect. 
The comparison confirms that dijets are produced in nuclear collisions down to 3 GeV jet energy (the most probable energy) but not  lower, and that fragments from low-energy jets extend down to 1 GeV/c (trigger) and 0.35 GeV/c (associated), the data consistent only with FFs from \ppbar\ collisions.

\subsection{Isolation of triggered dijets from secondaries}

In a given \pp\ event only the triggered dijet can be compared with the predictions from FFs that are predicated on a single dijet being present in $4\pi$. However, in \pp\ collisions there is a nonzero probability that secondary dijets may appear within the detector acceptance, and for higher multiplicities secondary dijets approach a certainty.
Based on \pp\ spectrum dijet systematics we can predict accurately the frequency of dijets in all \pp\ collision events and the number of dijets per hard event. We also predict the probability that a trigger hadron in a hard event is jet-related. We can then sort the full TA associated hard component into three parts: (a) soft trigger vs hard A (all dijets), (b) hard trigger vs secondary dijets and (c) hard trigger vs triggered dijet. Only combination (c) should be correlated and is the primary objective of this study. The other terms are modeled by the uncorrelated TCM hard component in Fig.~\ref{ppspec} (c) and subtracted, the required amplitudes obtained accurately from spectrum studies.

%\subsection{Implications for UE}

%\subsection{There really are lots of hadrons from jets}

%at low momenta

%%%%%%%%%%%%%%%%%%
\section{Summary}  \label{sum}

In previous work we reported trigger-associated (TA) correlations from 200 GeV \pp\ collisions as a joint distribution $F$ on trigger and associated hadron transverse rapidities $y_t$. The joint distribution can be factored into a trigger spectrum $T$ and associated distributions $A$ conditional on trigger rapidity. We defined a two-component (soft+hard) model (TCM) of TA correlations and isolated data hard components of $A$ that can be associated with dijet production. Specifically, hard component $A_{hh}$ includes jet correlations that are of primary interest.

In the present study we construct a quantitative prediction for TA hard component $A_{hh}$ derived from measured parton fragmentation functions (FFs) and a minimum-bias (MB) jet spectrum. The derivation transforms parton-fragment conditional distributions to hadron trigger-associated conditional distributions by integrating over the parton degree of freedom. The probability chain rule and Bayes' theorem are used in the transformation.

The comparison between predictions and data for four combinations of parton type (gluon or quark) and jet context (\ee\ or \ppbar\ collisions) indicates that only one combination (quark jets from \ppbar\ collisions) is consistent with all the data, but for that combination there is quantitative agreement at the 10\% level except at the kinematic boundaries of jet formation. Such comparisons provide another quantitative link between pQCD and minimum-bias jet structure in spectra and correlations. In the process kinematic lower bounds on jet formation are newly established separately for trigger and associated hadron fragments. This analysis  confirms that minimum-bias jets make a large contribution to \pp, \pa\ and \aa\ collisions especially at lower hadron momenta ($\approx 0.5$ - 1 GeV/c).

The analysis includes isolation of single triggered dijets from other untriggered secondary dijets that become prominent in higher-multiplicity \pp\ collisions. The secondary dijets can be identified with {\em multiple parton interactions} (MPI) that are a subject of underlying-event (UE) studies. Given the trigger-hadron direction the associated conditional distribution can be obtained for separate azimuth regions corresponding to those conventionally defined in UE studies. We observe a substantial contribution from the triggered dijet to the transverse or ``trans'' region symmetric about $\pi /2$, whereas the conventional assumption in UE analysis is that the triggered dijet makes no contribution to the trans region.

%%%%%%%%%%%%%%%%%%
This material is based upon work supported by the U.S.\ Department of Energy Office of Science, Office of Nuclear Physics under Award Number DE-FG02-97ER41020.

\end{document}